\documentclass[12pt]{article}
 \usepackage{graphicx,amsmath,amssymb}
\usepackage{indentfirst}
 \usepackage[usenames]{color}
 \usepackage[colorlinks=true, urlcolor=navyblue, linkcolor=navyblue, citecolor=navyblue]{hyperref}
\usepackage{epstopdf}
\usepackage{appendix}
\usepackage{enumerate}
\usepackage[symbol*]{footmisc}

\definecolor{navyblue}{rgb}{0,0.08,0.45}

\def\Dslash{\raise.15ex\hbox{/}\kern-.7em D}
\def\Pslash{\raise.15ex\hbox{/}\kern-.7em P}

\newcommand{\beq}{\begin{equation}}
\newcommand{\enq}{\end{equation}}
\newcommand{\beqa}{\begin{eqnarray}}
\newcommand{\beqast}{\begin{eqnarray*}}
\newcommand{\enqa}{\end{eqnarray}}
\newcommand{\enqast}{\end{eqnarray*}}
\newcommand{\beml}{\begin{multline}}
\newcommand{\enml}{\end{multline}}
\newcommand{\nn}{\nonumber}
\newcommand{\req}[1]{(\ref{#1})}

\newcommand{\pa}{\partial}

\newcommand{\bec}{\begin{center}}
\newcommand{\enc}{\end{center}}
\newcommand{\beqo}{\begin{quote}}
\newcommand{\enqo}{\end{quote}}

\newcommand{\half}{{\textstyle{\frac{1}{2}}}}

\newcommand{\cL}{{\cal L}}

\newcommand{\al}{\alpha}

\newcommand{\ep}{\epsilon}
\newcommand{\ze}{\zeta}

\newcommand{\la}{\lambda}

\newcommand{\vp}{\varphi}

\newcommand{\La}{\Lambda}

\begin{document}

\begin{flushright}
{
\small
SLAC--PUB--15894\\
\date{today}}
\end{flushright}

\vspace{60pt}

\centerline{\Large \bf Modified Anti-de-Sitter Metric, }

\vspace{10pt}

\centerline{\Large \bf Light-Front Quantized QCD,}

\vspace{10pt}

\centerline{\Large \bf  and Conformal Quantum Mechanics\footnote{Presented by HGD at the International
Fall Workshop {\sc \href{http://geometryandphysics2.gie.im}{Geometry and Physics II}}, Institut Henri Poincar\'e, Paris, November 28 and 29, 2013}}

\vspace{20pt}

\centerline{{
Hans G\"unter Dosch,$^{a}$ 
\footnote{E-mail: \href{mailto:gdt@asterix.crnet.cr}{dosch@thphys.uni-heidelberg.de}}
Stanley J. Brodsky$^{b}$ 
\footnote{E-mail: \href{mailto:sjbth@slac.stanford.edu}{sjbth@slac.stanford.edu}}
and
Guy F. de T\'eramond,$^{c}$ 
\footnote{E-mail: \href{mailto:gdt@asterix.crnet.cr}{gdt@asterix.crnet.cr}}
}}

\vspace{30pt}

{\centerline {$^{a}${\it Institut f\"ur Theoretische Physik, Philosophenweg 16, D-6900 Heidelberg, Germany}}

\vspace{4pt}

{\centerline {$^{b}${\it SLAC National Accelerator Laboratory, 
Stanford University, Stanford, CA 94309, USA}}

\vspace{4pt}

{\centerline {$^{c}${\it Universidad de Costa Rica, San Jos\'e, Costa Rica}}

 \vspace{60pt}

\begin{abstract}

We briefly review the remarkable connections between light-front QCD, gravity in AdS space, and conformal quantum mechanics.  We discuss, in particular, the group theoretical and geometrical aspects of the underlying one-dimensional quantum field theory.    The resulting effective theory leads to a phenomenologically successful confining interaction potential  in the  relativistic light-front wave equation which incorporates relevant  non-perturbative dynamical aspects of hadron physics.

\end{abstract}

\newpage

\section{Introduction}
\label{intro}

The title of this contribution,  {\sc Modified Anti-de-Sitter metric, Light-Front Quantized QCD, and Conformal Quantum mechanics} fits nicely into the general theme of this {\sc Geometry and Physics} conference;  however, it contains no highbrow mathematics and is very phenomenological. It is mainly based on a recent publication in
Physics Letters \cite{Brodsky:2013ar}.  The talk is organized into three sections:
\begin{enumerate}[1)]

\item Some crucial problems in the treatment of strong interactions.

\item A very superficial sketch of  an astonishing relation between classical gravity and a quantum field theory which appears to be relevant for strong interactions, and,
 
\item Some results obtained by combining elements from these different worlds.
\end{enumerate}

\section{Nonperturbative QCD}

It is generally believed that we know the underlying theory of the strong interactions, that is of protons, neutrons, pions, etc.  It is a quantum field theory which is  invariant under the gauged $SU(3)$ symmetry group, called Quantum  Chromodynamics (QCD). The fundamental fermion fields are the quark fields, which carry color quantum numbers, referring to the $SU(3)$ group. They interact via the gauge bosons of the theory, the gluons.  In many respects, the theory is similar to Quantum Electrodynamics (QED), the theory of electrons and photons, the gauge theory of $U(1)$.  In contrast  to electrons and photons, however,  quarks and gluons do not appear in the Fock space of observable particles; they are permanently confined within  the hadrons.

A  problem, common to all realistic relativistic quantum field theories,  is especially flagrant in QCD:  the only  known analytically tractable treatment is perturbation theory, which obviously is not the most practical tool for a strongly interacting theory with permanently bound constituents.  But even in weakly interacting theories, such as QED, there is a need for semiclassical equations in order to treat bound states. Atomic physics without the Dirac or Schr\"odinger equation would be in a rather desolate state. Therefore there is a formidable task in QCD: Find and justify a semiclassical approach!  This task is not completely hopeless for several reasons:
\begin{enumerate}[i)]

\item The  quark model, based mainly on a Schr\"odinger equation with relativistic corrections is qualitatively  astonishingly successful (See {\it e.g.} \cite{PDG:2012}, Sec. 14).

\item   There are striking regularities in the hadronic spectra, notably Regge trajectories, which show a linear relation between the squared mass and the intrinsic angular momentum of hadrons (See {\it e.g.} \cite{Donnachie:2002en}).

\item  If one implements  light-front (LF) quantization, one obtains a  Hamiltonian framework for treating bound states in  relativistic theories  based on front-form dynamics~\cite{Dirac:1949cp, Brodsky:1997de}. It is based not on initial conditions at equal times, $x^0 = 0$, but on  the light-cone null plane  $x^+ = x^0+ x^3 = 0$.  In  this framework one obtains an effective frame-independent eigenvalue equation for the  Fock state of  a meson consisting of two massless quarks~\cite{deTeramond:2008ht}:
 \beq \label{LF}
\left( -\frac{d^2}{d \zeta^2}  + \frac{4L^2-1}{4 \zeta^2} + U(\zeta)
\right) \psi(\zeta) =M^2 \psi(\zeta) ,
\enq 
where $\zeta^2 = b^2_\perp x(1-x)$ is the invariant separation of the quark and antiquark in the transverse (1-2)  light-front plane, $x= {k^+\over P^+} = {k^0+k^3\over P^0+ P^3}$ is the quark light-front momentum fraction,  and $L=L^3$ is the eigenvalue of the relative orbital angular momentum.  The eigenvalues of this equation are the squared hadron masses  $P^2_\mu = M^2$.
\end{enumerate}

\section{AdS/CFT Correspondence and Light-Front \\Holographic QCD}

The search for semiclassical equations obtained a strong advance some 15 years ago by the so called Maldacena Conjecture~\cite{Maldacena:1997re,Gubser:1998bc,Witten:1998qj} . Roughly speaking, it states that a quantum gauge field theory in 4 dimensions corresponds to a classical gravitational theory in 5 dimensions. The  generating functional of the quantum  gauge field theory  is given by the minimum of the  classical action of the gravitational theory at a 4-dimensional border of the 5-dimensional space. The gravitational theory is determined by the anti-de Sitter  (AdS) metric in a 5-dimensional space, AdS$_5$. In Poincar\'e coordinates $x^0,x^1, \dots z=x^5$, where the border to the physical space is given by $z=0$,  the line element is
 \beq \label{line} 
 ds^2 = \frac{R^2}{z^2} \left( \sum_{i=0}^3 dx_i\,dx^i - dz^2\right),
 \enq
where $R$ is the AdS radius.

In practice, there are several undesirable features in this correspondence, notably the 4-dimensional quantum field theory  is heavily over-symmetric: it is a {\bf conformal supersymmetric}  gauge theory. Therefore, for phenomenological purpose it is more promising to follow a bottom-up approach, that is to start from a  realistic 4-dimensional quantum field  theory and look for a corresponding higher dimensional classical non-Euclidean theory of which the realistic theory is the holographic picture. In this talk we shall concentrate on an approach called Light-Front Holographic QCD, which was developed by two of the authors~\cite{ deTeramond:2008ht, Brodsky:2006uqa}.

Consider a scalar field in AdS$_5$. The invariant action is given by the invariant integration over the 5-dimensional scalar expression of the Lagrangian
$\mathcal{L} = g^{MN} \pa_M \Phi(x,z) \pa_N \Phi(x,z) - \mu^2 \Phi^2(x,z)$
 \beq  \label{act} 
 S= \int d^4x\, dz\, \sqrt{|g|} \left(g^{MN} \pa_M \Phi(x,z) \pa_N \Phi(x,z) - \mu^2 \Phi^2(x,z)\right),
 \enq
where $\mu$ is the AdS mass, which is {\it a priory} an arbitrary parameter.

We are looking for a field, which at the border $z=0$ describes a free hadron with momentum $P$, that is $\Phi(x,z) = e^{i P \cdot x} \Phi(z)$.  In this case, the Euler-Lagrange equation from the action \req{act} can be brought into the form
 \beq \label{eom-ads}
\left(- \frac{d^2}{dz^2}  + \frac{4(\mu R)^2 +16 -1}{4 z^2}\right) \phi(z) = M^2 \phi(z).
 \enq
Comparing  this equation of motion with the semiclassical equation \req{LF} one observes the same structure if one identifies the  AdS  variable $z$ with the LF variable  $\zeta$ and  $(\mu R)^2+4$ with {$L^2$}.   The critical value  $L=0$  corresponds to the lowest possible stable solution for $P^2 \ge 0$, the ground state of the LF Hamiltonian, in agreement with the AdS stability bound   $(\mu R)^2 \ge - 4$~\cite{Breitenlohner:1982jf}.  There is, however, no interaction term in  \req{eom-ads}, that is  {$U(\zeta) = 0$}. This is not surprising:  AdS$_5$ is a maximally symmetric space with 15 isometries which induce in the border Minkowski space  the symmetry under the conformal group {\it Conf}$\,(R^{1,3}$) with 15 generators: 10 Poincar\'e transformations, 4 inversions, and 1 dilatation. This conformal symmetry implies that there can be no scale in the theory and therefore also no discrete spectrum. The only way out is to distort the maximal symmetry present in the action. This can be done most easily by inserting a so called dilaton term into the action, that is by the modification
\req{act} to
 \beq  \label{actdil}
S= \int d^4x\, dz\, \sqrt{|g|} \mbox{{$e^{\phi(z)} $}}\left(g^{MN}
\pa_M \Phi(x,z) \pa_N \Phi(x,z) - \mu^2 \Phi^2(x,z)\right).
 \enq
The equation of motion derived from this action yields a non-vanishing potential:
  \beq \label{pot} U(z) = \frac{1}{4}
(\vp'(z))^2 - \frac{3}{z} \vp'(z) + \frac{1}{2} \vp''(z).
 \enq
A phenomenologically successful choice is the ``soft-wall" model \cite{Karch:2006pv}, in which $\phi(z) = \la \, z^2$. It leads to the potential~\cite{deTeramond:2012rt,deTeramond:2013it}
 \beq \label{potsw} 
U(z) = \la^2 z^2  -2 \la.
\enq

The description of higher-spin states is a more complex task since the covariant derivatives in the action includes the affine connection and, in principle, one has also to take into account all possible permutations in the tensor indices for arbitrary spin $J$. Here again, one can take advantage of  the mapping of the higher-dimensional equations to the LF Hamiltonian equation   \req{LF}. This procedure allows a clear distinction between the kinematical and dynamical aspects of the problem. Accordingly, the non-trivial geometry of pure AdS space encodes the kinematics,  and the additional deformations of AdS encode the dynamics, including confinement ~\cite{deTeramond:2013it}, as well as determining the form of the LF effective potential. One finds~\cite{deTeramond:2013it, deTeramond:2010ge}
\beq \label{U}
U(\ze, J) = \frac{1}{2}\vp''(\ze) +\frac{1}{4} \vp'(\ze)^2  + \frac{2J - 3}{2 \zeta} \vp'(\ze) ,
\enq
provided that the product of the AdS mass $\mu$ and the AdS curvature radius $R$ are related to the total and orbital light-front angular momentum, $J$ and  $L$. The specific form of the dilaton profile   $\varphi(z) = \la z^2$  leads through \req{U} to the effective LF potential 
\beq \label{U2}
U(\ze, J) =   \la^2 \ze^2 + 2 \la (J - 1),
\enq
with eigenvalues
\beq
M_{n, J, L}^2 = 4 \la \left(n + \frac{J+L}{2} \right),
\enq
where $n$ is the radial excitation quantum number, leading to daughter trajectories.
To describe baryons, one  considers the propagation of Dirac fields for arbitrary half-integer spin (Rarita-Schwinger fields)  in AdS space and the corresponding mapping to light-front physics in physical space-time~\cite{deTeramond:2013it, Gutsche:2011vb}.

\begin{figure}[ht]
\centering
\includegraphics[width=7.0cm]{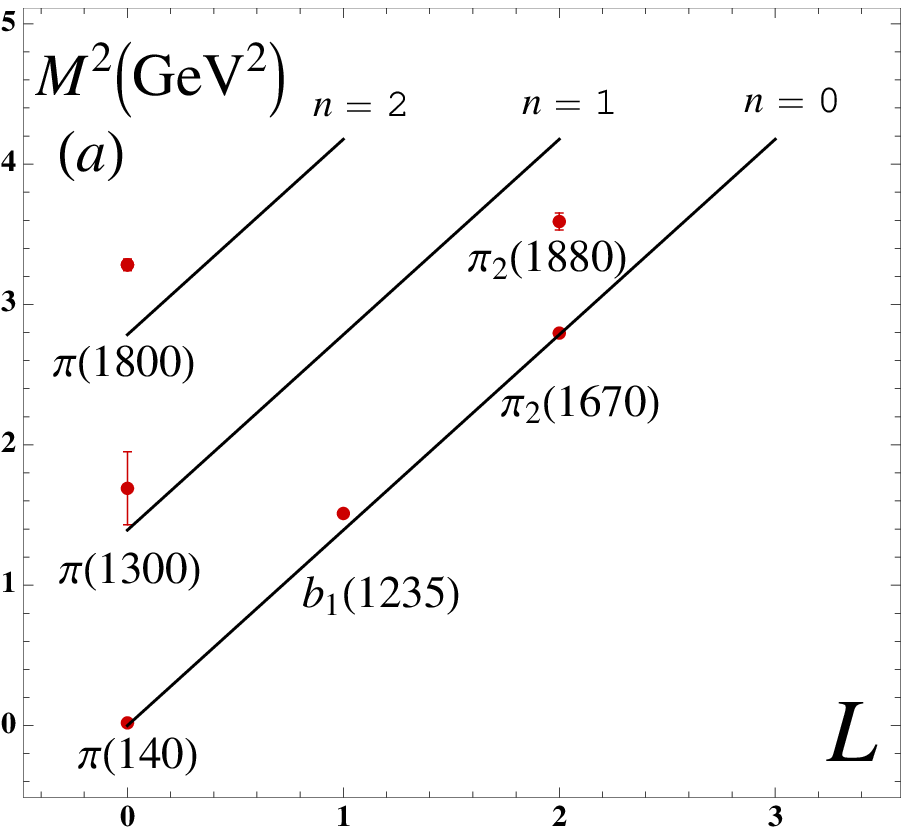}  \hspace{30pt}
\includegraphics[width=7.0cm]{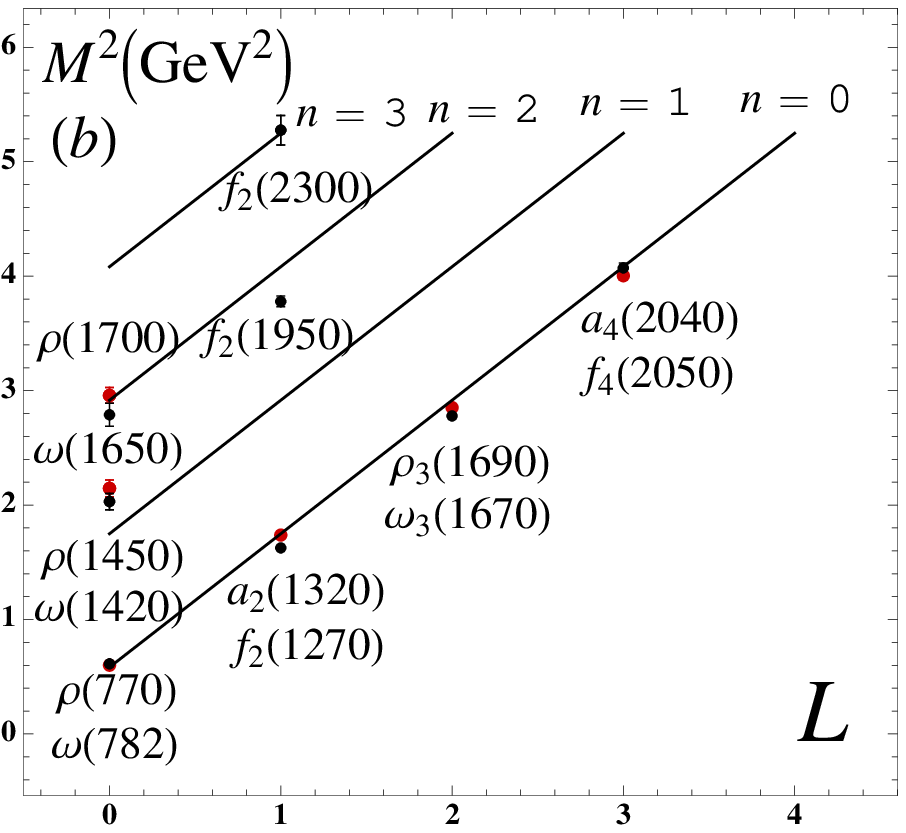}

 \vspace{20pt}
 
  \includegraphics[width=7.0cm]{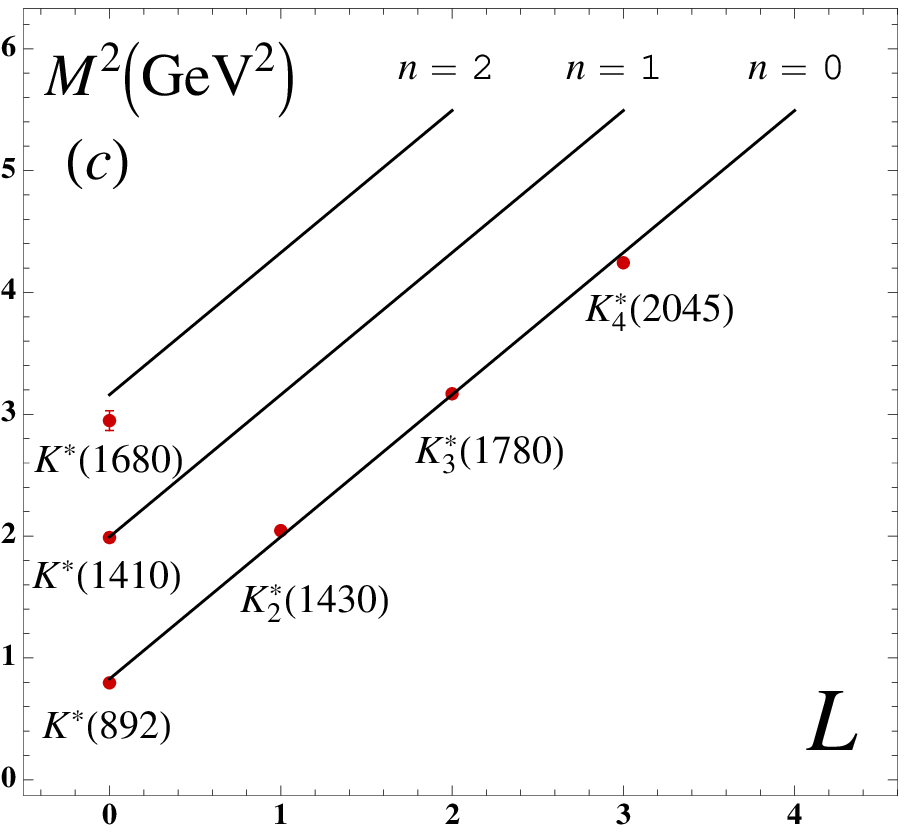}  \hspace{25pt}
 \includegraphics[width=7.0cm]{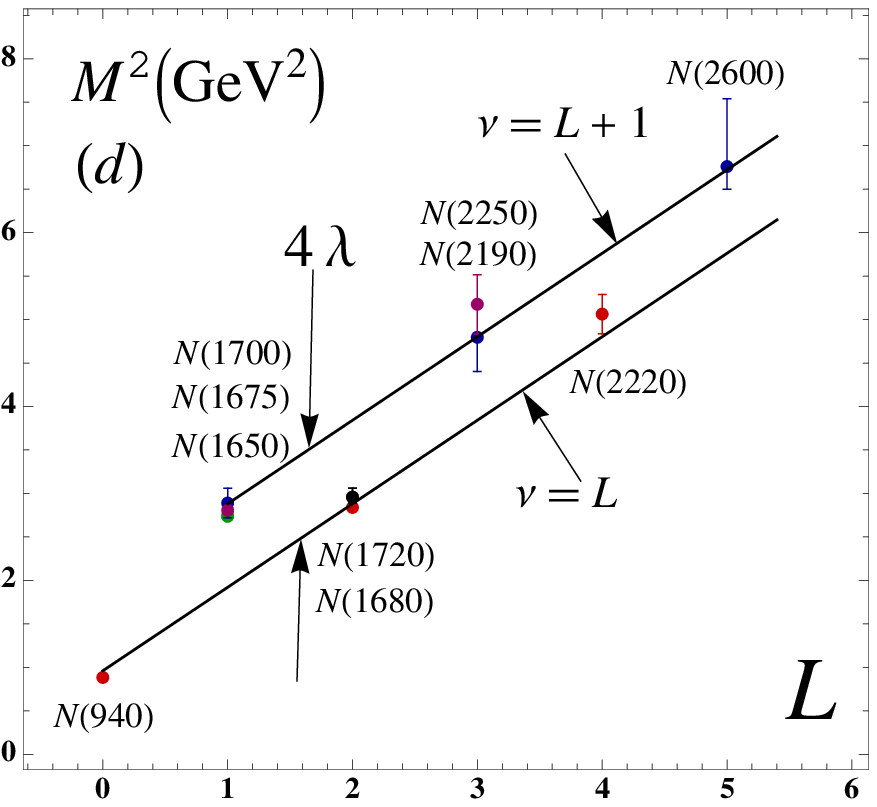}  
\caption{\label{fig-data}  Experimental values  and theoretical predictions for mesons and nucleons. For the non-strange mesons with isospin $I=1$ and with  internal spin $S=0$, {\it i.e.},   $C=(-1)^L$,  the optimal value is $\sqrt{\la}= 0.59$ GeV.  For mesons with $I = 0,1$ and $S=1$ and natural parity (non-strange and strange) $\sqrt{\la}= 0.54$ GeV, and for nucleons $\sqrt{\la}= 0.49$ GeV. Data are from \cite{PDG:2012}. Theoretical results from \cite{deTeramond:2012rt, deTeramond:2014yga}.}
\label{zerfall}
\end{figure}
This model yields linear Regge trajectories    with the same slope in the radial quantum number $n$ and orbital angular momentum $L$ as found experimentally.  A comparison with data is displayed in Fig.~\ref{fig-data}  for  light unflavored mesons and nucleon families. More details are given, for example, in Ref.~\cite{deTeramond:2014yga}.  The predictions can also be extended to other light hadron families, as for example the strange vector meson $K^*$ family which are also included in Fig. \ref{zerfall}. Good agreement prevails also, for example, in the model predictions for electromagnetic elastic and transition form factors~\cite{deTeramond:2012rt}. 

An unsatisfactory aspect, however,  is  that the specific  choice  $\varphi(z) = \la \, z^2$ is motivated only by phenomenology. One would like to derive it from some general principle. This is indeed possible as will be shown in Sect. \ref{CS} and \ref{GA}.

\section{Conformal Symmetry and its Consequences \label{CS}}

We now take a closer look at the implications of conformal symmetry. In QCD this symmetry plays a somewhat hidden role: The classical QCD-Lagrangian with massless quarks is conformally invariant, but this symmetry is broken due to quantum corrections. Indeed, the need for renormalization of the theory introduces a scale  $\La_{\rm QCD}$ which leads to the ``running coupling'' $\al_s(Q^2)$, (See {\it e.g.}  \cite{PDG:2012}, Sec. 9.1.1.)
 \beq
 Q^2 \, \frac {d \al_s(Q^2) }{d Q^2}  = -\sum_{i=0} b_i \, \al_s^{2+i} ,
 \enq
 with the solution
 \beq
\al_s(Q^2) = \frac{1}{b_0} \, \frac{1}{\log(Q^2/\La_{\rm QCD}^2)} + \cdots  .
\enq
The constants $b_i$ can be calculated in perturbation theory~\footnote{ The $b_i$ depend on the number of active flavours, for our case $ b_0 = 27/(12 \pi)$.},  but the so obtained
values are only reliable in the region where $\al_s$ is small, $\al_s \ll 1$, that is, for large values of $Q^2$, $Q^2 \gg \La_{\rm QCD}^2$.
There are, however, indications that at large distances, that is for small $Q$ values, $Q^2 < \La_{\rm QCD}^2$,  the coupling becomes constant again (See  \cite{Brodsky:2007hb} and literature quoted there). This  indicates a restoration of conformal symmetry in the non-perturbative regime in which we are interested.

Therefore we have a new aspect of conformal symmetry in QCD.  In our
approach  essential nonperturbative  aspects of a quantum field
theory are described  by a semiclassical equation, that is in a  quantum
mechanical description;  thus, we are motivated to investigate
conformal quantum mechanics, a  quantum field theory  in one
dimension, the time. It has been investigated thoroughly by
 V.~de Alfaro, S.~Fubini and G.~Furlan \cite{deAlfaro:1976je} some 37 years ago.

V.~de Alfaro {\it et al.},  start with the conformally invariant  action
   \beq \label{Sconf} S_{\it
conf}= \half \int dt\, \left(\dot Q(t)^2 -\frac{g}{Q(t)^2} \right) ,
\enq
 where $g$ is a dimensionless constant.
 The field momentum operator is  $P=\frac{\delta S}{\delta \dot Q} = \dot{Q}$, therefore quantization implies  $[Q,\dot Q]= i$ and the Hamiltonian is
 \beq \label{H}
H=\half\left(\dot Q^2 + \frac{g}{Q^2}\right) .
  \enq

We now go to the  Schr\"odinger picture in the state space of square integrable functions in the single variable $\psi(r) \in \cL_2(R^1)$.  We can represent $ Q(0) $ by the multiplication operator  $r$, and $\dot Q(0)$ by the differentiation operator $ - i \frac{d}{dr} $. This leads to the form of the Hamiltonian:
  \beq
H \psi(r) = \half\left( -  \frac{d^2}{ d r^2} + \frac{g}{r^2}\right) \psi({r}),
  \enq
and we are back again at the free case, \req{LF} with $U(\zeta)=0$, which also corresponds to the equation of motion \req{eom-ads} derived unmodified AdS$_5$. As mentioned above, this is not astonishing for a conformal theory. The dimensionless constant $g$ in  action \req{Sconf} is now  related to the Casimir operator of rotations  in the light front equation equation \req{LF}.    

However,  as stressed by de Alfaro, Fubini and Furlan~\cite{deAlfaro:1976je}, there are besides $H$, which is the generator of translations  in time $t$, two more constants of motion, namely the two Noether currents of the conformal action $S_{\it conf}$: $D$  for dilatations, $ t \to t(1 + \ep)$ and $K$ for special conformal transformation $t \to \frac{t}{1- \ep t}$.
This allows us to construct a generalized Hamiltonian: 
\beq  \label{G} 
G = H + w\, K + v\, D ,
\enq which describes a translation in a new ``time'' variable $\tau$ with
 \beq \label{tau}
d\tau= \frac{dt}{1+v t + w t^2} .
\enq

In the  Schr\"odinger picture $G$ reads:
  \beq
 G\; \psi(r)= \frac{1}{2} \left( -  \frac{d^2}{ d r^2} +  \frac{g}{r^2}   + \frac{i\,  v}{2} 
  \left( r \frac{d}{dr} + \frac{d}{dr}  r \right)  + w\, r^2\right)   \psi({r}).
  \enq
Identifying $r= \zeta/\sqrt{2} $ and $ g= L^2- 1/4$, we see that we get agreement with the light front Hamiltonian \req{LF} if we put $v=0$. In that case, the light front potential $U(\zeta)$ is
uniquely fixed to $U(\zeta)= w \,\zeta^2$. The confining Hamiltonian, \beq \label{Htau}  G = H + w \,K,
 \enq
 is, like $H$,   a translation operator, but not in the variable $t=x^0$, but in the variable $\tau = \frac{1}{\sqrt{w}}\, \arctan(\sqrt{w}\;t)$,  which has a finite range. Comparison with the equation of motion, derived from the distorted action \req{actdil}, fixes the dilaton profile to be quadratic in $z,  \; \vp(z) = w\, z^2$. This is exactly the form which leads to satisfactory agreement with the data.  The constant term of the potential \req{U2}, which is a kinematical consequence of  the AdS$_5$ action~\cite{deTeramond:2013it}, cannot be derived by these symmetry considerations.

\section{Geometrical Aspects \label{GA}}

The conformal group {\it Conf}$\,( R^1)$ is isomorphic to the Lorentz group $SO(2,1)$ and therefore also isomorphic to the isometries of AdS$_2$. This is best seen by embedding AdS$_2$ as an hyperboloid into  a 3-dimensional Euclidean  space with Cartesian coordinates $X_{-1},\,X_0,\,X_1$. In this case AdS$_2$ is the surface described by
 \beq \label{hyp}
X_{-1}^2 + X_0^2 - X_1^2= R^2.
 \enq
The Poincar\'e coordinates are related to the embedding coordinates by:
 \beq
z= \frac{R^2}{X_{-1}-X_1}, \hspace{30pt} 
x^0 = \frac{X_0 (X_{-1} - X_1)}{R}= X_0\, \frac{z}{R}.
 \enq
 \begin{figure}[ht]
\begin{center}
\includegraphics*[width=8cm]{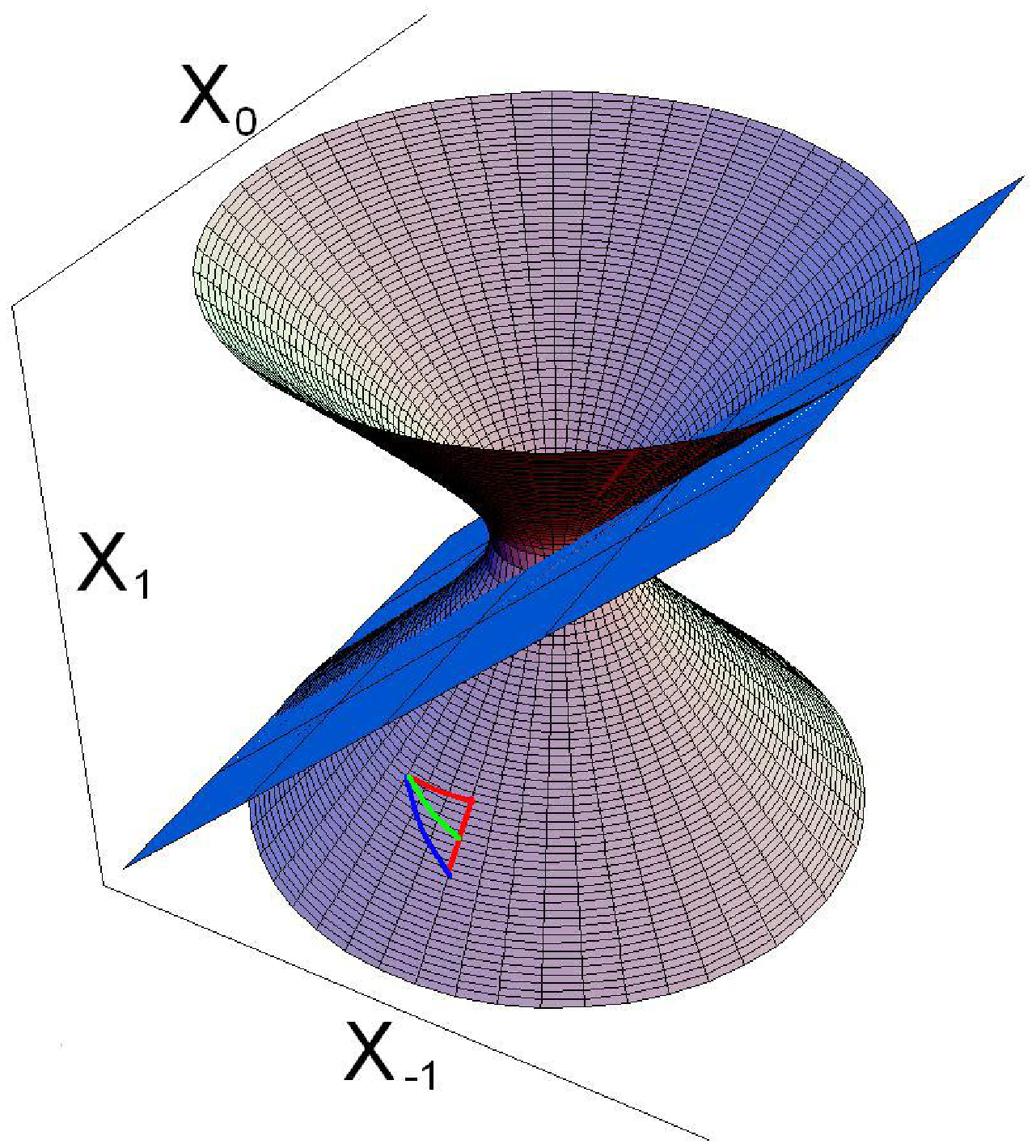}  \vspace{-0pt} 
\includegraphics*[width=4cm]{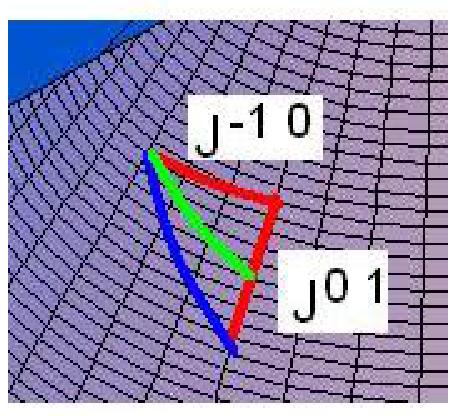}
\end{center}
\caption{\label{hypfig} The hyperboloid \req{hyp}, representing the non-Euclidean space AdS$_2$,  and the plane $X_1=X_{-1}$, which separates AdS$_2$ into two patches.  The red lines represent the infinitesimal transformation of the  boost $J^{-10}$ and the rotation $J^{01}$, the blue line of the Hamiltonian $H$ and the green line of the confining ``Hamiltonian'' $G$.  }
\end{figure}

 In Fig.~\ref{hypfig} the embedded rotation hyperboloid, representing the space AdS$_2$ and the plane $X_1 = X_-1$, which separates AdS$_2$ into two patches at $z=\pm \infty$  are displayed. The border $z=0$ is the intersection with the plane $X_{-1}-X_1 \to \infty$. The elements of $SO(2,1)$ are transformations on the hyperboloid. The generators of $SO(2,1)$ are the two boosts $J^{01}$ and $ J^{-11}$ in the $X_1$ direction,   and the rotation $J^{-1 0}$ in the  $(X_{-1},X_0)$ plane: they transform the hyperboloid into itself and are the isometries of  AdS$_2$. Due to the local isomorphism between $SO(2,1)$ and the conformal group  {\it Conf}$\,( R^1)$ we can relate the generators of the two groups. For the time translation operator $H$ \req{H}, the special conformal generator $K$ and the dilatation generator $D$ one obtains~\footnote{Since the generators of the conformal group have dimensions, a  constant $a$ with dimension dim[$a$] = - dim[$t$] occurs in these relations.}:
 \beqa
a\, H &=&  J^{-1 0}-J^{01} \label{comp}, \\
\frac{1}{a}K &=& J^{-1 0}+J^{01}, \nn \\
 D&=& J^{-1,1}. \nn
\enqa

The free Hamiltonian $H$ and the generator of the special conformal transformation $K$ are both linear combinations of the boost $J^{0 1}$  in $X_1$-direction and the rotation $J^{-10}$ in the  $(X_{-1},X_0)$ plane  \req{comp} (The infinitesimal action of the generator $H$ is depicted as the blue line in Fig.~\ref{hypfig}). Therefore, the confining Hamiltonian, $G$ \req{Htau}, can also be expressed  as linear combinations of these generators.
\beq
\frac{1+\theta}{2}\, a\,  G = J^{-1 0}- \theta\, J^{01}.
\enq
In the   Schr\"odinger picture it has thus the form
  \beq
\label{Htau}  \mbox{{$ G$}}= \half \left(- \frac{d^2}{d r^2} + \frac{g}{r^2} +
\frac{1}{a^2} \frac{1-\mbox{{$\theta$}}}{1+\mbox{{$\theta$}}}
r^2\right).
  \enq
This shows that, apart from a general scaling factor, the confining Hamiltonian $G$   can be viewed as a transformation in which the rotation and the boost are out of tune (Green line in Fig. \ref{hypfig}). The dimensionless coefficient $\theta$, which has to be  $\theta=1$  for the free Hamiltonian, can take any value $-1 < \theta < 1$ for the confining Hamiltonian.  Its numerical value is determined by $\La_{\rm QCD}$.

 It is interesting to note that a translation operator $G$ with $v\neq 0$, which is excluded in light front holographic QCD,  cannot be obtained in this way, since it contains the boost $J^{-1 1}$ which does not contribute to the free Hamiltonian $H$ (See \req{comp}). Therefore the modification from the free Hamiltonian $H$ \req{H}  to the confining Hamiltonian $G$ \req{Htau} is a sort of minimal modification.

\section{Conclusions}

To summarize: The combination of light-front quantized holographic QCD with symmetry considerations in conformal quantum mechanics yields a remarkably consistent and phenomenologically successful basis for establishing a semiclassical bound-state equation  for light hadrons in non-perturbative QCD.   The form of the interaction is uniquely fixed by the requirement of a minimal modification of the free Hamiltonian leaving the action invariant.

\section*{Acknowledgement}

We thank the organizers of this conference, in particular Joseph Kouneiher and  John Stachel, for the opportunity to give a seminar at this prestigious place. 
It is a pleasure to thank John Stachel also for his interest and valuable remarks about the Unimodular Group.


\begin{thebibliography}{}

  \bibitem{Brodsky:2013ar} 
  S.~J.~Brodsky, G.~F.~de Teramond and H.~G.~Dosch,
  ``Threefold complementary approach to holographic QCD,''
  \href{http://www.sciencedirect.com/science/article/pii/S0370269313010198}{Phys.\ Lett. \ B {\bf 729}, 3 (2014)}
 [\href{http://arxiv.org/abs/arXiv:1302.4105}{\tt arXiv:1302.4105 [hep-th]}].


\bibitem{PDG:2012}
 J.~Beringer {\it et al.} (Particle Data Group), ``Review of Particle Physics,'' \href{http://prd.aps.org/abstract/PRD/v86/i1/e010001}{Phys.\ Rev. \ D {\bf 86}, 010001 (2012).}


\bibitem{Donnachie:2002en}
  S.~Donnachie, H.~G.~Dosch, O.~Nachtmann and P.~Landshoff,
  ``Pomeron physics and QCD,''
  Camb.\ Monogr.\ Part.\ Phys.\ Nucl.\ Phys.\ Cosmol.\  {\bf 19}, 1 (2002).


 \bibitem{Dirac:1949cp}
  P.~A.~M.~Dirac,
  ``Forms of relativistic dynamics,''
 \href{http://rmp.aps.org/abstract/RMP/v21/i3/p392_1}{ Rev.\ Mod.\ Phys.\  {\bf 21}, 392 (1949)}.

 
 \bibitem{Brodsky:1997de}
  S.~J.~Brodsky, H.~C.~Pauli and S.~S.~Pinsky,
  ``Quantum chromodynamics and other field theories on the light cone,''
  \href{http://www.sciencedirect.com/science/article/pii/S0370157397000896}{Phys.\ Rept.\  {\bf 301}, 299 (1998)}
  [\href{http://arXiv.org/abs/hep-ph/9705477}{\tt arXiv:hep-ph/9705477}].
  
  
   \bibitem{deTeramond:2008ht}
  G.~F.~de Teramond and S.~J.~Brodsky,
  ``Light-front holography: a first approximation to QCD,''
  \href{http://prl.aps.org/abstract/PRL/v102/i8/e081601}{ Phys.\ Rev.\ Lett.\  {\bf 102}, 081601 (2009)}
[\href{http://arXiv.org/abs/0809.4899}{\tt arXiv:0809.4899 [hep-ph]}].

    
   \bibitem{Maldacena:1997re}
 J.~M.~Maldacena,
 ``The large N limit of superconformal field theories and supergravity,''
 \href{http://www.springerlink.com/content/q508214382421612/}{Int.\ J.\ Theor.\ Phys.\  {\bf 38}, 1113 (1999)}
 [\href{http://arXiv.org/abs/hep-th/9711200}{\tt arXiv:hep-th/9711200}].


   \bibitem{Gubser:1998bc}
  S.~S.~Gubser, I.~R.~Klebanov and A.~M.~Polyakov,
  ``Gauge theory correlators from non-critical string theory,''
  \href{http://www.sciencedirect.com/science/article/pii/S0370269398003773}{Phys.\ Lett.\ B {\bf 428}, 105 (1998)}
  [\href{http://arXiv.org/abs/hep-th/9802109}{\tt arXiv:hep-th/9802109}].


\bibitem{Witten:1998qj}
  E.~Witten,
  ``Anti-de Sitter space and holography,''
  Adv.\ Theor.\ Math.\ Phys.\  {\bf 2}, 253 (1998)
  [\href{http://arXiv.org/abs/hep-th/9802150}{\tt arXiv:hep-th/9802150}].
  
   
  \bibitem{Brodsky:2006uqa}
  S.~J.~Brodsky and G.~F.~de Teramond,
  ``Hadronic spectra and light-front wavefunctions in holographic QCD,''
  \href{http://prl.aps.org/abstract/PRL/v96/i20/e201601}{ Phys.\ Rev.\ Lett.\  {\bf 96}, 201601 (2006)}
  [\href{http://arXiv.org/abs/hep-ph/0602252}{\tt arXiv:hep-ph/0602252}].


  \bibitem{Breitenlohner:1982jf}
  P.~Breitenlohner and D.~Z.~Freedman,
  ``Stability in gauged extended supergravity,''
  \href{http://www.sciencedirect.com/science/article/pii/0003491682901166}{Annals Phys.\  {\bf 144}, 249 (1982)}.


\bibitem{Karch:2006pv}
  A.~Karch, E.~Katz, D.~T.~Son and M.~A.~Stephanov,
  ``Linear confinement and AdS/QCD,''
 \href{http://prd.aps.org/abstract/PRD/v74/i1/e015005}{ Phys.\ Rev.\  D {\bf 74}, 015005 (2006)}
  [\href{http://arXiv.org/abs/hep-ph/0602229}{\tt arXiv:hep-ph/0602229}].
  
  
     \bibitem{deTeramond:2012rt} 
  G.~F.~de Teramond and S.~J.~Brodsky,
 ``Hadronic form factor models and spectroscopy within the gauge/gravity correspondence,''
 \href{http://arxiv.org/abs/arXiv:1203.4025}{arXiv:1203.4025 [hep-ph]}. 
  
  
   \bibitem{deTeramond:2013it}
 G.~F.~de Teramond, H.~G.~Dosch and S.~J.~Brodsky,
 ``Kinematical and dynamical aspects of higher-spin bound-state equations in holographic QCD,''
 \href{http://prd.aps.org/abstract/PRD/v87/i7/e075005}{Phys.\ Rev.\ D {\bf 87}, 075005 (2013)}
 [\href{http://arxiv.org/abs/arXiv:1301.1651}{\tt arXiv:1301.1651 [hep-ph]}].
 
 
   \bibitem{deTeramond:2010ge}
G.~F.~de Teramond and S.~J.~Brodsky,
 ``Gauge/gravity duality and hadron physics at the light-front,''
\href{http://proceedings.aip.org/resource/2/apcpcs/1296/1/128_1?bypassSSO=1}{AIP Conf.\ Proc.\  {\bf 1296}, 128 (2010)}
[\href{http://arXiv.org/abs/1006.2431}{\tt arXiv:1006.2431 [hep-ph]}].
  
  
   \bibitem{Gutsche:2011vb}
  T.~Gutsche, V.~E.~Lyubovitskij, I.~Schmidt, A.~Vega,
  ``Dilaton in a soft-wall holographic approach to mesons and baryons,''
  \href{http://prd.aps.org/abstract/PRD/v85/i7/e076003}{ Phys.\ Rev.\ D {\bf 85}, 076003 (2012)}
  [\href{http://arXiv.org/abs/1108.0346}{\tt arXiv:1108.0346 [hep-ph]}].
  
  
  \bibitem{deTeramond:2014yga} 
  G.~F.~de Teramond, S.~J.~Brodsky and H.~G.~Dosch,
  ``Hadron spectroscopy and dynamics from light-front holography and conformal symmetry,''
  \href{http://arxiv.org/abs/arXiv:1401.5531}{\tt arXiv:1401.5531 [hep-ph]}.
  
 
  \bibitem{Brodsky:2007hb}
   S.~J.~Brodsky and G.~F.~de Teramond,
  ``Light-front dynamics and AdS/QCD correspondence: the pion form factor in the space- and timelike regions,''
  \href{http://prd.aps.org/abstract/PRD/v77/i5/e056007}{ Phys.\ Rev.\  D {\bf 77}, 056007 (2008)}
 [\href{http://arXiv.org/abs/0707.3859}{\tt arXiv:0707.3859 [hep-ph]}].


 \bibitem{deAlfaro:1976je}
  V.~de Alfaro, S.~Fubini and G.~Furlan,
  ``Conformal invariance in quantum mechanics,''
  \href{http://link.springer.com/article/10.10072FBF02785666}{Nuovo Cim.\ A {\bf 34}, 569 (1976)}.
  
 
  
 
\end{thebibliography}
\end{document}